\documentclass[a4paper,12pt]{article}
\pdfoutput=1

\makeatletter

\@addtoreset{equation}{section}
\makeatother
\usepackage{amsfonts}
\usepackage{amsmath}
\usepackage{latexsym}

\usepackage{mathrsfs}

\def\be{\begin{equation}}
\def\ee{\end{equation}}
\def\bea{\begin{eqnarray}}
\def\eea{\end{eqnarray}}
\def\({\left(}
\def\){\right)}
\def\<{\left<}
\def\>{\right>}

\def\[{\left[}
\def\]{\right]}

\def\be{\begin{equation}}
\def\ee{\end{equation}}
\def\bea{\begin{eqnarray}}
\def\eea{\end{eqnarray}}

\def\({\left(}
\def\){\right)}
\def\<{\left<}
\def\>{\right>}

\def\be{\begin{equation}}
\def\ee{\end{equation}}
\def\bea{\begin{eqnarray*}}
\def\eea{\end{eqnarray*}}
\def\ben{\begin{eqnarray}}
\def\een{\end{eqnarray}}
\def\({\left(}
\def\){\right)}
\def\<{\left<}
\def\>{\right>}
\def\!{\right|}
\def\|{\left|}

\def\[{\left[}
\def\]{\right]}

\def\+{\bar}
\def\mb{\mathbb}

\def\Vol{{\mbox{Vol}}}

\def\t{\widetilde}
\def\A{{\cal{A}}}

\def\N{{\cal{N}}}

\def\P{{\cal{P}}}

\def\H{{\mb{H}}}

\def\eps{{\cal{\varepsilon}}}

\def\l{{{\ell}}}

\def\h{\widehat}

\def\P{\mathscr{P}}
\def\N{\mathscr{N}}

\begin{document}

\pagestyle{empty}
\vskip-10pt
\vskip-10pt
\begin{center}
\vskip 3truecm
{\Large\bf
Abelian M5-brane on $S^6_q$}
\vskip 2truecm
{\large \bf
Andreas Gustavsson}
\vspace{1cm} 
\begin{center} 
Physics Department, University of Seoul, 13 Siripdae, Seoul 130-743 Korea
\end{center}
\vskip 0.7truecm
\begin{center}
(\tt agbrev@gmail.com)
\end{center}
\end{center}
\vskip 2truecm
{\abstract We compute the conformal anomaly of the abelian M5 brane on a conical deformation $S^6_q$ of the round six-sphere. Our results agree with corresponding results on $S^1 \times \mathbb{H}^5$ that were obtained in arXiv:1511.00313. For the free energies we obtain missing Casimir energy contributions, inconsequental for the Renyi entropies, and we obtain the proposed constant shift for the Renyi entropy of the selfdual two-form.}

\vfill
\vskip4pt
\eject
\pagestyle{plain}

\section{Introduction}
The $r$-dependence of the abelian partition function on $S^6$ was obtained in \cite{Gustavsson:2019efu} from the spectrum of various differential operators on $S^6$. This $r$-dependence of the partition function is governed by the M5 brane conformal anomaly that for abelian gauge group was obtained in \cite{Bastianelli:2000hi}. In this paper we will refine the computation in \cite{Gustavsson:2019efu} to a conically deformed six-sphere $S^6_q$, also called the branched sphere \cite{Nishioka:2013haa}. For integer values of $q = 1/\gamma$ we have $q$ different branches of $S^6$. 

One motivation for considering such a deformation is to introduce another continuous parameter $\gamma$ in addition to the radius $r$. As we have one more parameter, we may compute the partition function that will be on the form
\bea
Z(q,r) &=& c(q) r^{\A(q)}
\eea
In this paper we will only compute the conformal anomaly $\A(q)$.

Another motivation for considering $S^6_q$ is to study the gauge/gravity duality and in particular how entanglement entropy \cite{Ryu:2006bv}, \cite{Casini:2011kv}, and more generally Renyi entropy \cite{Huang:2014gca}, \cite{Zhou:2015kaj}, is mapped from gauge theory to gravity theory. This has been studied in the past literature by using a conformal map from $S^d_q$ to $S^1_q \times \H^{d-1}$ for various dimensions, following \cite{Casini:2011kv}. Despite there is a conformal anomaly, we can compute the conformal anomaly itself on either manifold and get the same answer. This is because by the Wess-Zumino consistency condition, the anomaly has no anomaly itself, essentially because $d^2 = 0$. When there is no conformal anomaly and $\A(q) = 0$, as is the case in odd dimensions, then we can still compute $c(q)$ and this will be conformally invariant simply because there is no conformal anomaly. This quantity was computed on the gauge theory side and for free field theories where an agreement was explicitly demonstrated for the free energies computed for both $S^3_q$ and $S^1_q \times \H^2$ \cite{Klebanov:2011uf}. 

In \cite{Nian:2015xky} both the nonsupersymmetric and the supersymmeric Renyi entropies are computed for the abelian M5 brane on $S^1_q \times \H^5$ and the generalization of the supersymmetric Renyi entropy to the nonabelian M5 brane was obtained in \cite{Zhou:2015kaj} and more generally to the nonabelian (1,0) SCFT's in \cite{Yankielowicz:2017xkf}.

The quantity that we are computing goes by several different names in the literature: the conformal anomaly $\A(q)$, the heat-kernel coefficient $a_6(q)$, the Minakshisundaram–Pleijel zeta function $\zeta(0;q)$ evaluated at $s=0$. In our convention they are all equal to one another, $\A(q) = a_6(q) = \zeta(0;q)$ and they are functions of $q$. If we ignore the prefactor $c(q)$, then it is also related to the one-loop effective action $W$ and to the free energy $F$, as $Z = r^{a_6} = e^{-W} = e^{-\beta F}$ where $\beta = 2\pi q$ is the inverse temperature. 

Our results for the nonsupersymmetric case for the conformal anomalies are 
\ben
a_6^S &=& \frac{1}{15120 q^5} + \frac{1}{4320q^3} + \frac{31q}{30240}\label{confscal}\\
a_6^B &=& \frac{6}{15120 q^5}+ \frac{1}{144q^3}+ \frac{1}{6q}   + \frac{31}{45} + \frac{191q}{1008}  \label{OurB}\\
a_6^F &=& - \frac{31}{120960 q^5}- \frac{7}{3456 q^3}- \frac{1}{128 q} -\frac{367 q}{24192}   \label{aF}
\een
for the conformal scalar, two-form gauge field and fermions, respectively. When we put $q = 1$, these results reproduce the results in \cite{Gustavsson:2019efu} for the round $S^6$,
\bea
a_6^S &=& \frac{1}{756}\cr
a_6^B &=& \frac{221}{210}\cr
a_6^F &=& - \frac{191}{7560}
\eea
The normalization of these conformal anomalies is such that for the 6d (2,0) tensor multiplet the conformal anomaly on $S^6$ is computed as
\bea
a_6^{(2,0)} = 5 a_6^S + \frac{1}{2} a_6^B - 2 a_6^F = \frac{7}{12}
\eea
Here $5$ is from the five conformal scalars on $S^6$, $1/2$ is because we work with a non-selfdual two-form, and the factor of $2$ corresponds to the  $2$ in the (2,0) tensor multiplet. On a generic smooth six-manifold (without conical singularities), the conformal anomaly receives contributions from the Weyl tensor and the structure of the conformal anomaly is 
\bea
\A &=& a_6 E_6 + c_1 I_1 + c_2 I_2 + c_3 I_3
\eea
where the explicit expressions for $E_6, I_1, I_2, I_3$ can be found in \cite{Bastianelli:2000hi}. In \cite{Beccaria:2017lcz} it was argued that the coefficient $c_3$ can be computed as 
\bea
c_3 &=& \frac{1}{12} \(a''_6 + 2 a'_6\)|_{q=1}
\eea
We can confirm this by computing this number for each field in the tensor multiplet, from (\ref{confscal}), (\ref{OurB}) and (\ref{aF}) respectively. We then get
\bea
c_3^S &=& \frac{1}{2520}\cr
c_3^B &=& \frac{1}{28}\cr
c_3^F &=& -\frac{1}{252}
\eea
The same coefficient $c_3^S$ was computed in \cite{Beccaria:2017lcz} with the same result. We now see that the ratios between these three numbers are in precise agreement with the ratios between the coefficients $c_3$ that were obtained in \cite{Bastianelli:2000hi} as $c_3^S:c_3^B:c_3^F = 2:180:40$. For the agreement, one has to take into account that their convention is such that $\A = 5 a_6^S + \frac{1}{2} a_6^B + a_6^F$, which differs by a factor of $-2$ for $a_6^F$ from our convention. So we see that although these coefficients are computed in a case when supersymmetry is completely broken when $q$ is deformed away from $q=1$, they apparently still contain some very useful information. 

The corresponding nonsupersymmetric results that were obtained in \cite{Nian:2015xky} on $S^1_q\times \H^5$ are\footnote{We extract these results from \cite{Nian:2015xky} by putting $\Vol(\H^5) = \pi^2 \ln(r)$ and $\beta = 2\pi q$ there.}
\ben
a_6^S &=& \frac{1}{15120 q^5} + \frac{1}{4320 q^3}\\
a_6^B &=& \frac{6}{15120 q^5} + \frac{1}{144 q^3}  + \frac{1}{6 q} \label{Nian}\\
a_6^F &=& - \frac{31}{120960 q^5} - \frac{7}{3456 q^3} - \frac{1}{128 q}
\een
Also in \cite{Nian:2015xky} the supersymmetric Renyi entropies were obtained.

Firstly, we see that results on $S^6_q$ differ from results on $S^1_q\times \H^5$ by a term proportional to $q$. In \cite{Beccaria:2017lcz} such a term was argued to correspond to the Casimir energy $E_C$ on $\mb{R} \times S^{d-1}$ and the following general structure was found for the conformal anomaly on $S^d_q$ for $d=4$ and $d=6$,
\bea
a_6 &=& - \frac{\nu}{3(d+1)! q^{d-1}} + ... + (- 2E_C) q
\eea
Indeed we find this structure for the conformal scalar and the two-form gauge field. For the fermion we do not find this structure. The reason is simply because we break supersymmetry and to preserve supersymmetry we need to turn on a background R-gauge field. Doing so, we obtain a supersymmetric $(\gamma-1)$-deformation of $a_6^F$ that reads
\ben
\t a_6^F &=& \frac{4}{15120 q^5} + \frac{1}{1080 q^3} - \frac{11}{360} + \frac{31 q}{7560}   \label{nice}
\een
The R gauge field does not couple to the two-form and one of the five conformal scalars so those conformal anomalies remain unchanged in the supersymmetric setting. The supersymmetric Casimir energies for the Abelian M5 brane on $\mb{R}\times S^5$ can be extracted from the following single particle indices \cite{Bak:2016vpi},
\bea
&&f_S(t) = \frac{e^{2t}(1+e^t)}{(e^t-1)^5} = \frac{2}{t^5} + ... + \frac{31}{30240} t\cr
&&f_{B^+}(t) = \frac{1-5e^t+10e^{2t}}{(e^t-1)^5} = \frac{6}{t^5} + ... + \frac{191}{2016} t\cr
&&f_F(t) = -\frac{4 e^{2t}(1+e^t)}{(e^t-1)^5} = - \frac{8}{t^5} + ... - \frac{31}{7560} t
\eea
for the conformal scalar, the two-form and the fermions respectively. The general structure of these single particle indices is 
\bea
f(t) &=& \frac{2\nu}{t^5} +... + (- 2E_C) t
\eea
where $\nu$ is the number of degrees of freedom and $E_C$ is the supersymmetric Casimir energy.

Secondly, there is a constant shift by $31/45$ for the two-form gauge field. In \cite{Nian:2015xky} a corresponding constant shift was proposed for the Renyi entropy. The Renyi entropy is \cite{Baez}\footnote{In \cite{Baez} the partition function is a function of temperatature, $\t Z(T)$. Here it is a function of $\beta = 1/T$. Thus $\t Z(T/q) = Z(q/T)=Z(q\beta)$.}
\bea
S_q &=& \frac{\ln Z(q\beta) - q \ln Z(\beta)}{1-q}
\eea
In this paper, we will suppress the overall factor $\ln(r)$ and thus define
\ben
S_q &=& \frac{a_6(q) - q a_6(1)}{1-q}\label{Renyi}
\een
We see from this expression that a Casimir energy term linear in $q$ does not contribute to $S_q$. A constant shift of $F(q)$ will contribute to $S_q$. More specifically, if $F(q)$ is shifted to $F(q) + C$, then $S_q$ gets shifted to $S_q + C$. We thus conclude that 
\bea
S^{S^6_q}_q &=& S^{S^1_q\times \H^5}_q + \frac{31}{45}
\eea
This is in exact agreement with the constant shift that was conjectured in \cite{Nian:2015xky}. 

In section \ref{Sq} we describe the geometry of $S^6_q$. In section \ref{Sp} we obtain the spectrum of two-form harmonics on $S^6_q$ after first reviewing the scalar and vector harmonics following \cite{DeNardo:1996kp}. In section \ref{BMN} we compute $a_6$ for the two-form gauge potential. In section \ref{CSc} we compute $a_6$ for a conformal scalar. In section \ref{Rep} we present a representation theory method to compute $a_6$. First we reproduce the results for scalar and vector harmonics by this method, and then we use this method to compute $a_6$ for one fermion. In section \ref{susyresults} we consider the supersymmetric case. In section \ref{disc} we discuss results in the previous literature in relatation to our results. 

There are five appendices. In section \ref{Box} we study the box operator on $S^6$ for higher rank differential forms. In section \ref{Lap} we study the laplace operator. In section \ref{factorization} we show how one can compute $a_6$ from a `half heat kernel' if the eigenvalues factorize following \cite{Beccaria:2017lcz}. In section \ref{Fer} we obtain solutions of the Dirac operator on $S^6$, although we do not compute degeneracies. In section \ref{S1} we discuss the solution of the Killing spinor equation on $S^1_q$ and its generalization to $S^6_q$.

\section{Note added}
After the publication of this article, I was informed that conformal anomalies on branched spheres had been previously obtained by Dowker \cite{Dowker:2017qkx},
\cite{Dowker:2017flz} for the nonsupersymmetric case. In particular, my result for the conformal anomaly of the two-form gauge field in eq (\ref{OurB}) had been previously obtained in \cite{Dowker:2017flz}. 

In addition, Dowker informed me that it is not always true that a determinant factorizes even if the eigenvalues factorize. There may be a factorization anomaly and my derivation in the appendix \ref{factorization}, which assumes the determinant factorizes, is then not correct. Nevertheless, Dowker assured me that this method works when used for the computation of anomalies in this paper due to some fortunate cancelation when one computes the average.

\section{The geometry of $S^6_q$}\label{Sq}
The geometry of a conically deformed sphere is described in \cite{DeNardo:1996kp}. We begin by the conically deformed flat space $\mb{R}^{d+1}$ into which $S^d$ may be embedded. Let us denote the Euclidean coordinates on $\mb{R}^{d+1}$ by $x^I$ for $I=1,...,d,d+1$ and let
\bea
x^{\pm} &=& \frac{1}{\sqrt{2}} \(x^d \pm i x^{d+1}\)
\eea
such that $x^I = (x^i,x^+,x^-)$ where $i=1,...,d-1$. The Euclidean flat metric on $\mb{R}^{d+1}$ is given by
\bea
ds^2 = dx^I dx^I = dx^i dx^i+2dx^+ dx^-
\eea
The cone $\mb{R}^7_{q}$ is now defined by making the identifications
\bea
x^{\pm} &\sim & e^{\pm 2\pi i q} x^\pm 
\eea
When $q=1,2,3,...$ this is a multi-cover of $\mb{R}^7$ and otherwise it is a cone. We also introduce the quantity
\bea
\gamma &=& \frac{1}{q}
\eea
The conically deformed sphere $S^d_{q}$ is the round sphere 
\bea
x^i x^i + x^+ x^- &=& r^2
\eea
embedded in $\mb{R}^{d+1}_{q}$. There is a conical singularity in $\mb{R}^{d+1}_q$ at $x^+ = x^- = 0$. This conical singularity is a submanifold $\mb{R}^{d-1}$ with coordinates $x^i$. It induces conical singularities on $S^d_q$ at $x^+ = x^- = 0$ where $x^i x^i = r^2$. Hence this conical singularity is an equatorial sphere $S^{d-2}$. Local coordinates on $S^d$ may be chosen such that its embedding in $\mb{R}^7$ is given by
\bea
x^i &=& r n^i \cos \theta \cr
x^d &=& r \sin \theta \cos \tau\cr
x^{d+1} &=& r \sin \theta \sin \tau
\eea
where $n^i n^i = 1$. Then the induced metric on $S^d$ is
\ben
ds^2 &=& r^2 \(d\theta^2 + \sin^2 \theta d\tau^2 + \cos^2 \theta d\Omega_{d-2}\)\label{induced}
\een
where $d\Omega_{d-2} = dn^i dn^i$ is the metric on unit $S^{d-2}$. Here $\tau \sim \tau + 2\pi$ and $\theta \in [0,\pi/2]$. The conical deformation is obtained by just changing the global identification to $\tau \sim \tau + 2\pi q$. This means that locally this is still $S^d$ away from $\theta = 0$. For $d=2$ there are two distinct conical singularities at the north and south poles at $\theta = 0$ for $n^1 = \pm 1$ respectively. For $d>2$ the conical singularity is not a point but an equatorial sphere $S^{d-2}$. The local geometry in the vicinity of this conical singularity is $\mb{R}^2_q \times S^{d-2}$. 

Let us now define a new coordinate $\theta_D = \pi/2 - \theta \in [0,\pi/2]$. The metric (\ref{induced}) becomes
\bea
ds^2 &=& r^2 \(d\theta_D^2 + \cos^2 \theta_D d\tau^2 + \sin^2 \theta_D d\Omega_{d-2}\)
\eea 
Expanding this metric to first order in $\theta_D$ around $\theta_D = 0$, we get
\bea
ds^2 &=& dt^2 + r^2 d\Omega_{d-1}
\eea
where $d\Omega_{d-1} = d\theta_D^2 + \sin^2 \theta_D d\Omega_{d-2}$ and $t = r \tau$. In other words, near the equator $x^i = 0$ we have locally the geometry of $S^1_q \times S^{d-1}$. In the limit $q\rightarrow \infty$ we have locally the geometry of $\mb{R} \times S^{d-1}$. 

In this paper we find that the asymptotic behavior of $a_6$ as $q \rightarrow \infty$ is governed by the supersymmetric Casimir energy $E_C$ that one computes on $\mb{R}\times S^{d-1}$. That one shall take this limit to see the Casimir energy is natural since the infinite $\beta$ limit of the partition function behaves in this limit as $Z \sim  e^{-\beta E_C}$. However, it is not immediately clear why it would be sufficient that only the local geometry near the equator is on the form $\mb{R}\times S^{d-1}$. On the other hand, $S^d_q$ in the infinite $q$ limit is conformally equivalent to $\mb{R} \times \H^{d-1}$. An explicit conformal transformation is presented between the two spaces for $d=3$ in \cite{Huang:2014gca}, which also shows that the tip of the cone of $S^d_q$ is mapped to the boundary of $\H^{d-1}$. If the supersymmetric Casimir energy is conformally invariant, as suggested by the fact that it appears in the conformal anomaly on $S^6_q$, then we should be able to compute it on $\mb{R} \times \H^{d-1}$ and get the same result as one gets on $\mb{R} \times S^{d-1}$. As far as we aware, the computation of the Casimir energy on $\mb{R} \times \H^{d-1}$ has not yet been done in the literature, see for instance \cite{Casini:2010kt}.

\section{Spectrum on $S^6_q$}\label{Sp}
We will now obtain the eigenvalues and degeneracies for scalar harmonics and vector harmonics on $S^6_q$. This will reproduce the result in \cite{DeNardo:1996kp}. We will then follow the same strategy to obtain a new result, the eigenvalues and degeneracies for two-form harmonics on $S^6_q$. 

Let us begin by the scalar harmonics on $S^d$. The scalar spherical harmonics are represented in $\mb{R}^{d+1}$ as a linear combination of elements of the form
\bea
\t\phi &=& \frac{1}{r^n} x^{I_1} \cdots x^{I_n} C_{I_1\cdots I_n}
\eea
for $n=0,1,...$. They are constructed such that we have $\partial_r \t\phi = 0$. Let us put $
\t\phi = \frac{1}{r^n} \phi$. Then we find that for $\square = \partial_I \partial_I$, 
\bea
\square \t\phi = \square \(\frac{1}{r^n}\) \phi + 2 \partial_I \(\frac{1}{r^n}\) \partial_I \phi + \square \phi
\eea
By using 
\bea
\square \(\frac{1}{r^n}\) &=& \frac{n(n+2)-n(d+1)}{r^{n+2}}\cr
\partial_I \(\frac{1}{r^n}\) \partial_I \phi &=& - \frac{n^2}{r^{n+2}} \phi
\eea
we get
\bea
\square \t\phi = -\frac{1}{r^2} n(n+d-1) \t\phi
\eea
provided we also assume that $\square \phi = 0$ which implies the tracelessness condition
\bea
\delta^{I_1I_2}C_{I_1I_2I_3\cdots I_n} &=& 0
\eea
The eigenvalue of $\square$ is entirely determined by the power of the prefactor $1/r^n$.

\subsection{Basis for harmonic differential forms}
We introduce the following basis elements, which are independent of $r$, on $S^6_q$,
\bea
\P^{\xi}_{n,m} &=& \frac{1}{r^{n+\gamma m}} (x^+)^{\gamma m + \xi} \sum_{p=0}^{\[\frac{n-\xi}{2}\]} x^{i_1} \cdots x^{i_{n-2p-\xi}} (x^+ x^-)^p (\P^{\xi})^{m}_{i_1\cdots i_{n-2p-\xi}}\cr
\N^{\xi}_{n,m} &=& \frac{1}{r^{n+\gamma m}} (x^-)^{\gamma m - \xi} \sum_{p=0}^{\[\frac{n+\xi}{2}\]} x^{i_1} \cdots x^{i_{n-2p+\xi}} (x^+ x^-)^p (\N^{\xi})^{m}_{i_1\cdots i_{n-2p+\xi}}
\eea
where $\xi \in \{-1,0,1\}$ gives the charge $Q = \xi q$ under the $U(1)_q$ rotation under which 
\bea
x^{\pm} &\rightarrow & e^{\pm 2\pi i q} x^{\pm}
\eea
Here $(\P^{\xi})^{m}_{i_1\cdots i_{n-2p-\xi}}$ are constant coefficients symmetric in the $i$ indices and subject to a tracelessness condition and some other regularity condition to be presented shortly. These basis elements provide a natural generalization of the scalar harmonics on $S^6$ to describe harmonic differential forms on $S^6_q$. A scalar field on $S^6_q$ carries charge $Q = 0$ as it shall be periodic. A vector field has components $v^I = (v^i, v^+,v^-)$ where $v^i$ carries charge $Q=0$ and $v^\pm$ carries charge $Q = \pm q$. The vector field components are expanded in the basis elements $(\P^i)_{n,m}, (\P^\pm)_{n,m}$ and $(\N^i)_{n,m}, (\N^\pm)_{n,m}$. Similarly, a two form has components $B^{IJ} = (B^{ij},B^{+-},B^{\pm i})$.

We want the basis elements to be regular everywhere. This means we need to avoid the presence of $(x^+)^{-1}$ and $(x^-)^{-1}$ respectively. Therefore we need to impose the following regularity constraints
\ben
(\P^{(-1)})^0_{i_1\cdots i_{n+1}} &=& 0\cr
(\N^{(+1)})^0_{i_1\cdots i_{n+1}} &=& 0\label{miszero}
\een
In addition, for vector and higher rank tensor harmonics, we need to assume that 
\bea
\gamma m - 1 \geq 0
\eea
for $m=1,2,...$ as one can see by demanding regularity of $\P^{-1}_{0,m}$. This means we need to assume that $q\leq 1$. One may then analytically continue to get results valid for $q > 1$.\footnote{One may embed $\mb{R}^2_q$ into $\mb{R}^3$ by cutting out a wedge with deficit angle $\alpha \in [0,2\pi]$ in a paper and gluing together the edges of the wedge. This produces a conical rice hat with $\beta = 2\pi - \alpha \leq 2\pi$. Two get $\beta = 2\pi + \alpha > 2\pi$ one can instead insert a wedge with deficit angle $\alpha \in [0,2\pi)$ into a half-slit that one has cut in a paper. One then gets something that takes the shape of a saddle. When $\beta = 4\pi$ one has a double cover of $\mb{R}^2$. One can keep inserting wedges indefinitely to create multicovered conical figures whose only singular point is the tip of the cone embedded into $\mb{R}^3$, but one will need to use several papers that one glues together along wedges.}

One can show that\footnote{Here it is being understood that the coefficients of $\P$ and $\N$ are being identified. Alternatively we may consider the class of all possible coefficients and then the equalities really mean equalities between classes of elements.} 
\bea
\P^{(0)}_{n,0} &=& \N^{(0)}_{n,0}\cr
\P^{(+1)}_{n,0} &=& \N^{(+1)}_{n,0}\cr
\P^{(-1)}_{n,0} &=& \N^{(-1)}_{n,0}
\eea
On the other hand, when $m=m'\neq 0$, we have $\P^{\xi}_{n,m'} \neq \N^{\xi}_{n,m'}$.

\subsection{The tracelessness constraint}
The computation of the eigenvalues for the scalar harmonics on $S^6$ can be repeated verbatim for the basis elements on $S^6_q$. The eigenvalues of $\square$ are fully determined by the power of the prefactor $1/r^{n+\gamma m}$ so we get
\bea
\square (\P^{\xi})_{n,m} &=& - \frac{1}{r^2} (n+\gamma m)(n+\gamma m+d-1) (\P^{\xi})_{n,m}
\eea
and we need to impose
\ben
\square \(r^{n+\gamma m}(\P^{\xi})_{n,m}\) &=& 0\label{traceless}
\een
In particular, the eigenvalues are not sensitive to the charge $\xi$.

Explicitly (\ref{traceless}) amount to the tracelessness conditions
\bea
(n-2p+2-\xi)(n-2p+1-\xi) (\P^\xi)^m_{iii_1\cdots i_{n-2p-\xi}} + 2 p (\gamma m+p+\xi) (\P^{\xi})^m_{i_1\cdots i_{n-2p-\xi}} &=& 0
\eea
These conditions are nontrivial for $p = 1,...,\[\frac{n-\xi}{2}\]$. By taking $p=0$ we get 
\bea
(\P^\xi)^m_{iii_1\cdots i_{n-\xi}} &=& 0
\eea
which are trivially realized since the highest nonvanishing components are $(\P^{\xi})^{m}_{i_1\cdots i_{n-\xi}}$. We will be mostly interested in the highest equation, that is $p=1$,
\ben
(n-\xi)(n-1-\xi) (\P^\xi)^m_{iii_1\cdots i_{n-2-\xi}} + 2 (\gamma m+1+\xi) (\P^{\xi})^m_{i_1\cdots i_{n-2-\xi}} &=& 0\label{trless}
\een
since the lower tracelessness constraints (those with $p>1$) are relating the all the lower components to the highest components. Thus to obtain the degeneracies, we only need to study the most general solutions to the highest equations. 

As an illustration of a general idea, let us consider the scalar harmonics. The highest components are $\P^m_{i_1\cdots i_n}$ and $\N^m_{i_1\cdots i_n}$ and all the lower components $\P^m_{i_1\cdots i_{n-2p}}$ and $\N^m_{i_1\cdots i_{n-2p}}$ are related to the highest components via the tracelessness constraints. The highest component $\P^m_{i_1\cdots i_n}$ is a rank $n$ symmetric tensor. As such it has 
\bea
d_{n,0} &=& \(\begin{array}{c}
n+d-2\\
n
\end{array}\)
\eea
independent components, where $d=6$ for $S^6_q$. When $m\neq 0$ there is an independent set of components $\N^m_{i_1\cdots i_n}$ so in total we have $d_{n,m\neq 0} = 2 d_{n,0}$, whereas when $m=0$ we have $d_{n,0}$ components. For the case of a round sphere, $\gamma = 1$, we have $p =n+\gamma m = n + m$ is an integer, and we can define a degeneracy 
\bea
d_p &=& \sum_{n+m=p} d_{n,m}
\eea
for scalar harmonics associated to the eigenvalue $p(p+d-1)$. It can be computed using the hockey-stick identity for binomial coefficients with the result 
\bea
d_p &=&  \(\begin{array}{c}
d+n\\
n
\end{array}\) - \(\begin{array}{c}
d+n-2\\
n-2
\end{array}\)
\eea
which corresponds to the number of components of a rank-$n$ traceless symmetric tensor in $d+1$ dimensions. This is a nice consistency check for the dimensions $d_{n,m}$.

\subsection{Transversality and coclosed constraints}
To descend to one-form harmonics on $S^d$, we shall impose the constraints
\bea
x^I v_I &=& 0\cr
\partial^I v_I &=& 0
\eea
In polar coordinates, the first condition amounts to the transversality constraint\footnote{One way to see this is by using $\partial_r x^I = \frac{x^I}{r}$.}
\bea
v_r &=& 0
\eea
while the second constraint becomes
\bea
D^m v_m &=& 0
\eea
where $\theta^m$ denote local coordinates on $S^6$. This is just saying that $v_m$ is coexact. 

The transversality and coexact constraints generalize to higher rank differential forms in a natural way,
\bea
x^I v_{II_1\cdots I_{p-1}} &=& 0\cr
\partial^I v_{II_1\cdots I_{p-1}} &=& 0
\eea

We now list the explicit form of the transversality and coexact constraints for the one-form and two-form harmonics.

\subsubsection{One-form}
\bea
(\P_{i_{(n-2p+1}})^{m}_{i_1\cdots i_{n-2p)}} + (\P^+)^{m}_{i_1\cdots i_{n-2p+1}} + (\P^-)^{m}_{i_1\cdots i_{n-2p+1}} &=& 0\cr
(n-2p) (\P_k)^m_{ki_1\cdots i_{n-2p-1}} + (\gamma m+p+1) (\P^+)^m_{i_1\cdots i_{n-2p-1}} + (p+1) (\P^-)^m_{i_1\cdots i_{n-2p-1}} &=& 0
\eea

\subsubsection{Two-form}
\bea
(\P_{(k}{}^j)^m_{i_1\cdots i_{n-2p})} + (\P^{+j})^m_{i_1\cdots i_{n-2p} k} + (\P^{-j})^m_{i_1\cdots i_{n-2p} k} &=& 0\cr
(\P_{(k}{}^+)^m_{i_1\cdots i_{n-2p-1})}-(\P^{+-})^m_{i_1\cdots i_{n-2p-1}k} &=& 0\cr
(\P_{(k}{}^-)^m_{i_1\cdots i_{n-2p-1})} + (\P^{+-})^m_{i_1\cdots i_{n-2p-1}k} &=& 0\cr
(n-2p) (\P^{ik})^m_{ii_1\cdots i_{n-2p-1}} + (\gamma m+1+p) (\P^{+k})^m_{i_1\cdots i_{n-2p-1}} + (p+1) (\P^{-k})^m_{i_1\cdots i_{n-2p-1}} &=& 0\cr
(\gamma m+p) (\P^{+-})^m_{i_1\cdots i_{n-2p}} + (n-2p+1) (\P^{i-})^m_{ii_1\cdots i_{n-2p}} &=& 0\cr
(p+1) (\P^{+-})^m_{i_1\cdots i_{n-2p-2}} - (n-2p-1) (\P^{i+})^m_{ii_1\cdots i_{n-2p-2}} &=& 0
\eea

\subsection{The highest equations}
The highest equations are obtained by taking $p=0$. We shall also note that the highest nonvanishing components are $(\P^{\xi})^m_{i_1\cdots i_{n-\xi}}$ unless $m=0$ in which case we have the constraints (\ref{miszero}), so that $(\P^{(-1)})^0_{i_1\cdots i_{n-1}}$ becomes the highest component with $\xi = -1$.

\subsubsection{One-form}
\ben
(\P_{i_{(n+1}})^{m}_{i_1\cdots i_{n)}} + (\P^-)^{m}_{i_1\cdots i_{n+1}} &=& 0\label{v1}\\
n (\P_k)^m_{ki_1\cdots i_{n-1}} + (\gamma m+1) (\P^+)^m_{i_1\cdots i_{n-1}} + (\P^-)^m_{i_1\cdots i_{n-1}} &=& 0\label{v2}
\een
When $m\neq 0$, all components are related to $(\P^k)^m_{i_1\cdots i_n}$. When $m=0$, the equation (\ref{v1}) gives rise to the constraints
\ben
(\P_{i_{(n+1}})^{0}_{i_1\cdots i_{n)}} &=& 0\label{c1v}
\een
and then we need to instead look at the next level, that is at $p=1$, where we find the constraints
\ben
(\P_{(i_{n-1}})^0_{i_1\cdots i_{n-2}} + (\P^+)^0_{i_1\cdots i_{n-1}} + (\P^-)^0_{i_1\cdots i_{n-1}} &=& 0\label{v1a}
\een
Let us now move on to (\ref{v2}). When $m=0$, this equation becomes
\bea
n (\P_k)^0_{ki_1\cdots i_{n-1}} + (\P^+)^0_{i_1\cdots i_{n-1}} + (\P^-)^0_{i_1\cdots i_{n-1}} &=& 0
\eea
and by using (\ref{v1a}), we get
\bea
(\P_{(i_{n-1}})^0_{i_1\cdots i_{n-2})} &=& n (\P_k)^0_{ki_1\cdots i_{n-1}}
\eea
We now notice that the tracelessness condition amounts to
\bea
2(\P_{i_{n-1}})^0_{i_1\cdots i_{n-2}} + n(n-1) (\P_{(i_{n-1}})^0_{i_1\cdots i_{n-2})ii} &=& 0
\eea
and so by combining these two conditions, we seem to get yet another constraint 
\ben
2 (\P_k)^0_{ki_1\cdots i_{n-1}} + (n-1) (\P_{(i_{n-1}})^0_{i_1\cdots i_{n-2}kk} &=& 0\label{c1vprime}
\een
But this is really not a new constraint, since (\ref{c1vprime}) follows beautifully from (\ref{c1v}) upon contraction of two indices,
\bea
\delta^{i_1 i_2} (\P_{(i_1})^0_{i_2\cdots i_{n})} &=& 0
\eea
Thus when $m=0$ the independent components are $(\P^-)^0_{i_1\cdots i_{n-1}}$ and $(\P^k)^0_{i_1\cdots i_n}$ subject to the constraints (\ref{c1v}). This leads to the degeneracy
\ben
d_{n,m=0} &=& \(\begin{array}{c}
d+n-3\\
d-2
\end{array}\) + (d-1) \(\begin{array}{c}
d+n-2\\
d-2
\end{array}\) - \(\begin{array}{c}
d+n-1\\
d-2
\end{array}\)\label{vectordeg0}
\een
When $m\neq 0$ the independent components are $(\P^k)_{i_1\cdots i_n}$ and $(\N^k)_{i_1\cdots i_n}$, leading to 
\ben
d_{n,m\neq 0} &=& 2(d-1) \(\begin{array}{c}
n+d-2\\
n
\end{array}\)\label{vectordeg1}
\een
These results were first obtained in \cite{DeNardo:1996kp}. When $q = 1$ we have the undeformed $S^6$ and we get
\bea
v_n &=& d_{n,0} + 2 \sum_{m=0}^{n-1} d_{m,n-m}
\eea
Explicit computation gives
\bea
v_n &=& \frac{n(n+d-1)(2n+d-1)(n+d-3)!}{(d-2)!(n+1)!}
\eea
which for $d=6$ reduces to
\bea
v_n &=& \frac{1}{24} n (n+2)(n+3)(n+5)(2n+5)
\eea

\subsubsection{Two-form}
\ben
(\P_{(k}{}^j)^m_{i_1\cdots i_{n})} + (\P^{-j})^m_{i_1\cdots i_{n} k} &=& 0\label{c1t}\\
(\P^+{}_{(k})^m_{i_1\cdots i_{n-1})} + (\P^{+-})^m_{i_1\cdots i_{n-1}k} &=& 0\label{c2t}\\
(\P^-{}_{(k})^m_{i_1\cdots i_{n-1})} - (\P^{+-})^m_{i_1\cdots i_{n-1}k} &=& 0\label{c3t}\\
n (\P^{ik})^m_{ii_1\cdots i_{n-1}} + (\gamma m+1) (\P^{+k})^m_{i_1\cdots i_{n-1}} + (\P^{-k})^m_{i_1\cdots i_{n-1}} &=& 0\label{c4t}\\
\gamma m (\P^{+-})^m_{i_1\cdots i_{n}} - (n+1) (\P^{-i})^m_{ii_1\cdots i_{n}} &=& 0\label{c5t}\\
(\P^{+-})^m_{i_1\cdots i_{n-2}} + (n-1) (\P^{+i})^m_{ii_1\cdots i_{n-2}} &=& 0\label{c6t}
\een

We can eliminate $\P^{+-}$,
\ben
(\P_{(k}{}^j)^m_{i_1\cdots i_{n})} + (\P^{-j})^m_{i_1\cdots i_{n} k} &=& 0\label{c1t}\\
(\P^+{}_{(k})^m_{i_1\cdots i_{n-1})} + (\P^-{}_{(k})^m_{i_1\cdots i_{n-1})} &=& 0\label{c23t}\\
n (\P^{ik})^m_{ii_1\cdots i_{n-1}} + (\gamma m+1) (\P^{+k})^m_{i_1\cdots i_{n-1}} + (\P^{-k})^m_{i_1\cdots i_{n-1}} &=& 0\label{c4t}\\
\gamma m (\P^-{}_{(i_1})^m_{i_2\cdots i_n)} - (n+1) (\P^-{}_i)^m_{ii_1\cdots i_n} &=& 0\label{515}\\
(2+\gamma m) (\P^+{}_i)^m_{ii_1\cdots i_{n-2}} + \frac{n-2}{2} (\P^+{}_{(i_1})^m_{i_2\cdots i_{n-2})ii} &=& 0\label{cpt}
\een
Here (\ref{cpt}) can be simplified to
\ben
(n-1) (\P^+{}_i)_{ii_1\cdots i_{n-2}}^m - (\P^+{}_{(i_1})^m_{i_2\cdots i_{n-2})} &=& 0\label{cpt1}
\een
by applying the tracelessness condition (\ref{trless}).

When $m\neq 0$, we may use (\ref{c1t}) to eliminate $\P^{-j}$,
\bea
(\P^{-j})^m_{i_1\cdots i_{n+1}} &=& - (\P_{(i_{n+1}}{}^j)^m_{i_1\cdots i_n)}\cr
(\P^{-j})^m_{i_1\cdots i_{n-1}} &=& \frac{n(n-1)}{2\gamma m} (\P_{(i_1}{}^j)_{i_2\cdots i_{n-1})ii} + \frac{n}{\gamma m} (\P_i{}^j)_{ii_1\cdots i_{n-1}}
\eea
Using this, we can show that (\ref{515}) gets automatically solved! 

We also see that (\ref{c23t}) gets automatically solved, essentially because $(\P_{(i_1 k})^m_{i_2\cdots i_{n-1})ii} = 0$ due to antisymmetry in $i_1 k$. 

So all that remains to check is (\ref{cpt1}) once we have solved for $\P^{+k}$ from (\ref{c4t}),
\bea
(\P^{+k})^m_{i_1\cdots i_{n-1}} &=&  - \frac{n(n-1)}{2(\gamma m+1) \gamma m} (\P_{(i_1}{}^k)^m_{i_2\cdots i_{n-1})ii} - \frac{n}{\gamma m} (\P_i{}^k)^m_{ii_1\cdots i_{n-1}}
\eea
that is further simplified to
\bea
(\P^{+k})^m_{i_1\cdots i_{n-1}} &=& \frac{1}{\gamma m} (\P_{(i_1}{}^k)^m_{i_2\cdots i_{n-1})} - \frac{n}{\gamma m} (\P_i{}^k)^m_{ii_1\cdots i_{n-1}}
\eea
by applying the tracelessness condition (\ref{trless}). 

Then finally we need to see whether (\ref{cpt1}) is also satisfied. This is a hard problem. Let us simply assume this is the case. 

Let us now summarize. For $m\neq 0$, we have found that the independent components are $(\P_{ij})^m_{i_1\cdots i_n}$. All the other tensors, as well as all the lower rank tensors are related to this one by the tracelessness conditions or by the conditions we have shown above. 

Let us now assume that $m=0$. When $m=0$, the equation (\ref{c1t}) implies the constraints
\ben
(\P_{(k}{}^j)^0_{i_1\cdots i_{n})} &=& 0\label{csn}
\een
We eliminate $\P^{+-}$,
\bea
(\P^+{}_{(k})^0_{i_1\cdots i_{n-1})} + (\P^-{}_{(k})^0_{i_1\cdots i_{n-1})} &=& 0\label{sum}\\
n (\P^{ik})^0_{ii_1\cdots i_{n-1}} + (\P^{+k})^0_{i_1\cdots i_{n-1}} + (\P^{-k})^0_{i_1\cdots i_{n-1}} &=& 0\label{mus}\\
(\P^{-i})^0_{ii_1\cdots i_n} &=& 0\label{auto}\\
(n-2) (\P^+{}_i)^0_{ii_1\cdots i_{n-2}} - (\P^+{}_{(i_1})^0_{i_2\cdots i_{n-2})} &=& 0\label{tricky}
\eea
where (\ref{tricky}) derives from (\ref{c6t}). We notice that (\ref{sum}) is implied by (\ref{mus}) and (\ref{csn}). Equation (\ref{auto}) is implied by (\ref{miszero}). 

We can now summarize our result. The independent components are given by $(\P^{ij})^0_{i_1\cdots i_n}$ and $(\P^{+i})^0_{i_1\cdots i_{n-1}}$ subject to the constraints (\ref{csn}) and (\ref{tricky}). The number of indepedent such components is given by
\bea
d^T_{n,m=0} &=& \frac{(d-1)(d-2)}{2} \(\begin{array}{c}
d+n-2\\
n
\end{array}\) - \frac{(d-1)(d-2)}{n+2} \(\begin{array}{c}
d+n-1\\
n
\end{array}\)\cr
&& + (d-1) \(\begin{array}{c}
d+n-3\\
n-1
\end{array}\) - \(\begin{array}{c}
d+n-4\\
n-2
\end{array}\)
\eea
This expression can be simplified to 
\bea
d^T_{n,0} &=& \frac{(d+n-2)(n(d-1)+d^2-6d+13)}{2(n+2)} \(\begin{array}{c}
d+n-4\\
n-1
\end{array}\)
\eea

For $d=6$ we get
\bea
d^T_{n,0} &=& \frac{1}{12} n (n+1)(n+4)(5n+13)
\eea

As for the case when $m\neq 0$, we have assumed that 
\bea
d^T_{n,m\neq 0} &=& (d-1)(d-2) \(\begin{array}{c}
d+n-2\\
n
\end{array}\)
\eea
So when $d=6$, we get
\bea
d^T_{n,m\neq 0} &=& \frac{5}{6} (n+1)(n+2)(n+3)(n+4)
\eea
As a consistency check, we then get 
\bea
d^T_{n,0} + \sum_{p=0}^{n-1} d^T_{p,m\neq 0} &=& t_n
\eea
where
\bea
t_n &=& \frac{1}{12} n (n+1)(n+4)(n+5)(2n+5)
\eea
is the degeneracy of two-form harmonics on round $S^6$.

\section{The two-form conformal anomaly}\label{BMN}
Let us now recall how we compute the heat kernel of a two-form gauge potential on round $S^6$. The spherical harmonics correspond to coexact forms. Therefore the heat kernel is computed as
\bea
a_6^B &=& a_6^T - a_6^V + a_6^{S_0}
\eea
up to some zero mode. Here $T$, $V$ and $S_0$ refers to the two-form, the two one-form ghosts, and the three massless scalar ghosts. The reason we do not have the coefficients two and three multiplying these ghost contributions here, is because we are working with the coexact parts only. For a more detailed explanation we refer to \cite{Bak:2016vpi}. To compute this, we thus need the refined dimensions for scalar, vector and two-form harmonics, which we obtained in the previous section. We also need the correponding eigenvalues. By using properties of the box operator $\square$ that we derive in appendix \ref{Box}, we find that when acting on massless scalar, vector and two-form harmonics respectively, we get
\bea
\t\square (S_0)_{n,m} &=& - (n+\gamma m)(n+\gamma m+d-1) (S_0)_{n,m}\cr
\t\square V_{n,m} &=& - \[(n+\gamma m)(n+\gamma m+d-1)-1\] V_{n,m}\cr
\t\square T_{n,m} &=& - \[(n+\gamma m)(n+\gamma m+d-1)-2\] T_{n,m}
\eea
Then by using the following relations that we derive in appendix \ref{Lap} between the laplacians acting on a scalar, one-form and two-form and the box operator
\bea
\t\triangle_0 &=& - \t\square\cr
\t\triangle_1 &=& - \t\square + 5\cr
\t\triangle_2 &=& - \t\square + 8
\eea
we get
\bea
\t\triangle_0 (S_0)_{n,m} &=& (n+\gamma m)(n+\gamma m+d-1) (S_0)_{n,m}\cr
\t\triangle_1 V_{n,m} &=& \[(n+\gamma m)(n+\gamma m+d-1)+4\] V_{n,m}\cr
\t\triangle_2 T_{n,m} &=& \[(n+\gamma m)(n+\gamma m+d-1)+6\] T_{n,m}
\eea
The determinant factorizes if the eigenvalues factorize. If we put $d=6$, then the factorized form of the eigenvalues is 
\bea
\t\triangle_p &=& (n+\gamma m+p)(n+\gamma m+5-p)
\eea
for $p=0,1,2$. As explained in \cite{Beccaria:2017lcz} and in appendix \ref{factorization}, each factor gives rise to its own `half heat kernel', 
\bea
K_{p}(t) &=& \sum_{n=0}^{\infty} \(d_{n,0} + d_{n,m\neq 0} \sum_{m=1}^{\infty} e^{-t\gamma m}\) e^{-t (n+p)}\cr
K_{5-p}(t) &=& \sum_{n=0}^{\infty} \(d_{n,0} + d_{n,m\neq 0} \sum_{m=1}^{\infty} e^{-t\gamma m}\) e^{-t (n+5-p)}
\eea
To extract the conformal anomaly $a_6$, we expand $(K_p+K_{5-p})/2$ around small $t$ and extract its constant term. We get
\ben
a_6^{S_0} &=& \frac{863}{6098\gamma} + \frac{\gamma}{6} - \frac{7\gamma^3}{864} + \frac{\gamma^5}{15120}\\
a_6^V &=& \frac{61}{90} - \frac{221}{6048 \gamma} - \frac{11\gamma^3}{864} + \frac{\gamma^5}{3024}\label{vektor}\\
a_6^T &=& - \frac{19}{30} + \frac{31}{3024\gamma} + \frac{\gamma^3}{432} + \frac{\gamma^5}{1512}
\een
and then 
\bea
a_6^B = a_6^T-a_6^V-a_6^{S_0} + 2 = \frac{31}{45} + \frac{191}{1008 \gamma} + \frac{\gamma}{6} + \frac{\gamma^3}{144} + \frac{\gamma^5}{2520}
\eea
We need to add $2$ by hand that accounts for two overcounted vector ghost zero modes  by the heat kernels \cite{Gustavsson:2019efu}, \cite{Christensen:1979iy}, \cite{Fradkin:1983mq}, \cite{Tseytlin:2013fca}.

\section{The conformal scalar}\label{CSc}
For the conformal scalar, we have eigenvalues 
\bea
\t\triangle_0 + 6 = (n+\gamma m)(n+\gamma m+5) + 6 = (n+\gamma m + 2) (n+\gamma m+3)
\eea
We then get  
\ben
a_6^{S} &=& \frac{31}{30240\gamma} + \frac{\gamma^3}{4320} + \frac{\gamma^5}{15120}\label{ConfScal}
\een
This result was first obtained in \cite{Beccaria:2017lcz}.

\section{Spectrum from representation theory}\label{Rep}
For the fermions we can not straightforwardly apply the same method as we did for they $p$-forms. We may try to solve the eigenvalue problem $\Gamma^M D_M \psi = \lambda \psi$ explicitly and find the eigenvalues and degeneracies by generalizing the approach in \cite{Klebanov:2011uf}. 

We will proceed in a different way and use $SO(7)$ representation theory. We obtain a universal method to compute the conformal anomaly on $S^6_q$ that seems to work for any field. We will use this universal method to reproduce the results that we got above for the scalar harmonics and vector harmonics. We then apply this method on the fermions. 

We will obtained branching rules for irreducible representations corresponding to scalar harmonics, vector harmonics and fermion harmonics respectively, under $SO(7) \rightarrow SO(5) \times U(1)_q$. Here $SO(7)$ is the isometry group of $S^6$, and $SO(5) \times U(1)_q$ is the isometry of  $S^6_q$. We then refine the dimension of the $SO(7)$ representation by turning on a chemical potential for the $U(1)_q$, but we do this in a novel way where instead of using the $U(1)_q$ charge for the refined dimension, we shall take the absolute value of the $U(1)_q$ charge. We explain why this prescription is needed for the scalar harmonic. We verify that this prescription also gives the right answer for the vector harmonics, which makes us rather confident that our method is universal and we then finally apply this method on the fermions. 

The representations of $SO(7)$ that we will need for the tensor multiplet on $S^6$ can be found in the appendix of \cite{Gustavsson:2019efu}. Let us here summarize the representations of $SO(5)$. We have the simple roots $\alpha_1$ and $\alpha_2$, the Cartan matrix
\bea
A_{ij} &=& \(\begin{array}{cc}
2 & -2\\
-1 & 2
\end{array}\)
\eea
We have the positive roots $\alpha_1,\alpha_2,\alpha_1+\alpha_2,\alpha_1+2\alpha_2$. Their sum is 
\bea
2 \delta &=& 3 \alpha_1 + 4 \alpha_2
\eea
The dimension of $SO(5)$ representation with Dynkin labels $(\Lambda_1,\Lambda_2)$ is 
\bea
\dim (\Lambda_1,\Lambda_2) &=& \frac{1}{6} (\Lambda_1+1)(\Lambda_2+1)(\Lambda_1+\Lambda_2+2)(2\Lambda_1+\Lambda_2+3)
\eea

\subsection{Scalar harmonics}
Scalar spherical harmonics on $S^6$ make up the irreducible representation $(n,0,0)$ of $SO(7)$. Under $SO(7) \rightarrow SO(5) \times U(1)$, we have the branching rule
\ben
(n,0,0) &\rightarrow & \bigoplus_{p=0}^n \bigoplus_{q=0}^p (n-p,0)_{p-2q}\label{branchscalar}
\een
For $n=1$ this amounts to $(1,0,0) \rightarrow (1,0)_0 \oplus (0,0)_{+1} \oplus (0,0)_{-1}$ which corresponds to the decomposition of a vector as $v^I = (v^i,v^+,v^-)$. One may verify that the dimensions add up correctly,
\bea
s_n := \dim (n,0,0) = \sum_{p=0}^n \sum_{q=0}^p \dim (n-p,0)
\eea
Here the dimension $s_n$ can be computed as the dimension of a rank $n$ symmetric tensor by removing traces,
\bea
s_n = \(\begin{array}{c}
n+6\\
n
\end{array}\) - \(\begin{array}{c}
n+4\\
n-2
\end{array}\) = \frac{1}{5!} (n+1)(n+2)(n+3)(n+4)(2n+5)
\eea
Let us now count the number of states in the representation $(n,0,0)$ with a given charge $Q = p - 2q$. If $Q=0$, then these states are in the representation
\bea
R_{n,0} &=& (n,0) \oplus (n-2,0) \oplus ... \oplus (n-2[n/2],0)
\eea
whose dimension is 
\bea
S_{n,0} = \sum_{q=0}^{[n/2]} \dim (n-2q,0) = \(\begin{array}{c}
n+4\\
4
\end{array}\)
\eea
Let us next find the representation with charge $Q = +1$. It is
\bea
R_{n,+1} &=& (n-1,0) \oplus (n-3,0) \oplus ... \oplus (n-1-2[(n-1)/2],0)
\eea
There is a corresponding representation $R_{n,-1}$ with charge $Q = -1$. The dimensions of each of these are
\bea
S_{n,+1} = \dim R_{n,+1} = S_{n-1,0}\cr
S_{n,-1} = \dim R_{n,-1} = S_{n-1,0}
\eea
In general, states with charge $Q = 0,\pm 1,\pm 2,...,\pm n$ are in a representation $R_{n,Q}$ of dimension $S_{n,Q} = \dim R_0 = S_{n,0}$ if $Q=0$ and $S_{n,Q}+S_{n,-Q} = \dim \(R_{+Q} \oplus R_{-Q}\) = 2 S_{n-Q,0}$ if $Q = 1,2,...,n$. 

Let us now consider the following refined heat kernel,
\bea
K_a(t) = \sum_{n=0}^{\infty} \sum_{Q=0}^n S_{n,Q} e^{-t(n+a)} e^{\alpha t Q} = \sum_{n=0}^{\infty} \(S_{n,0} + 2 \sum_{Q=1}^n S_{n-Q,0} e^{\alpha t Q}\) e^{-t(n+a)} 
\eea
Let us now compare this with 
\bea
K_a(t) = \sum_{n=0}^{\infty} \sum_{m=0}^{\infty} d_{n,m} e^{-t(n+\gamma m)} = \sum_{n=0}^{\infty} \(1 + 2 \sum_{m=1}^\infty e^{-t\gamma m}\) S_{n,0} e^{-t (n+a)} \cr
= \sum_{n=0}^{\infty} \(S_{n,0} + 2 \sum_{m=1}^n S_{n-m,0} e^{-t(\gamma-1)m}\) e^{-t (n+a)}
\eea
where we used $n+\gamma m = n+m + (\gamma -1)m$ and redefined $n+m$ as a new $n$ to go from the first to the second equality. By equating the two expressions, we find that 
\bea
\alpha &=& 1-\gamma\cr
Q &=& m
\eea
Finally let us contrast with the naturally refined the degeneracy
\bea
s_n(\alpha) := \sum_{p=0}^n \sum_{q=0}^p \dim(n-p,0) e^{t \alpha (p-2q)} = S_{n,0} + \sum_{m=1}^n S_{n-m,0} \(e^{\alpha t m} + e^{-\alpha t m}\)
\eea
This is not the one that we shall use. Instead we shall use the following refined degeneracy
\bea
s_n(\alpha) := \sum_{p=0}^n \sum_{q=0}^p \dim(n-p,0) e^{t \alpha |p-2q|} = S_{n,0} + 2 \sum_{m=1}^n S_{n-m,0} e^{\alpha t m} 
\eea
in the refined heat kernel 
\bea
K_a(t) &=& \sum_{n=0}^{\infty} s_n(1-\gamma) e^{-t (n+a)}
\eea

\subsection{Vector harmonics}
The vector harmonics appear in the product of a vector with scalar harmonics,
\bea
(1,0,0) \otimes (n,0,0) &=& (n+1,0,0) \oplus (n-1,0,0) \oplus (n-1,1,0)
\eea
The branching rule under $SO(7) \rightarrow SO(5) \times U(1)$ can be deduced from 
\bea
(1,0,0) &\rightarrow & (1,0)_0 \oplus (0,0)_{+1} \oplus (0,0)_{-1}\cr
(n,0,0) &\rightarrow & \bigoplus_{p=0}^n \bigoplus_{q=0}^p (n-p,0)_{p-2q}
\eea
and 
\bea
(n,0) \otimes (1,0) &=& (n+1,0) \oplus (n-1,0) \oplus (n-1,2)
\eea
From this, we can derive the branching rule
\bea
(n-1,1,0) &\rightarrow & \bigoplus_{p=0}^n \bigoplus_{q=0}^p (n-p-1,2)_{p-2q} \cr
&&\oplus \bigoplus_{p=0}^{n-1} \bigoplus_{q=0}^p \((n-p,0)_{p-2q+1}\oplus (n-p,0)_{p-2q-1}\)\cr
&&\oplus \bigoplus_{q=0}^{n-1} (0,0)_{n-1-2q}
\eea
One may check that the dimensions match,
\bea
\dim(n-1,1,0) &=& \sum_{p=0}^n (p+1) \dim(n-p-1,2) + 2\sum_{p=0}^{n-1} (p+1)\dim(n-p,0) + n \dim(0,0)
\eea
and that both sides amount to 
\bea
v_n = \dim(n-1,1,0) = \frac{1}{24} n(n+2)(n+3)(n+5)(2n+5)
\eea
The refined dimension is defined as
\bea
v_n(\alpha) &=& \sum_{p=0}^n \dim(n-p-1,2) \sum_{q=0}^p e^{\alpha t|p-2q|} \cr
&& + \sum_{p=0}^{n-1} \dim(n-p,0) \sum_{q=0}^p\(e^{\alpha t|p-2q+1|} + e^{\alpha t|p-2q-1|}\) \cr
&& + \sum_{q=0}^{n-1} e^{\alpha t|n-1-2q|}
\eea
Let us introduce the heat kernel  
\bea
K_{\alpha}(a,t) &=& \sum_{n=0}^{\infty} v_n(\alpha) e^{-t (n+a)}
\eea
Explicitly this is 
\ben
K_{\alpha}(a,t) &=& \sum_{\l=0}^{\infty} e^{-t(2\l+a)} \(\sum_{k=0}^{\l} \dim(2\l-2k-1,2) f(2k) + \sum_{k=0}^{\l-1} \dim(2\l-2k-2,2) f(2k+1)\)\cr
&+& \sum_{\l=0}^{\infty} e^{-t(2\l+1+a)} \(\sum_{k=0}^{\l} \dim(2\l-2k-1,2) f(2k+1) + \sum_{k=0}^{\l} \dim(2\l-2k-1,2) f(2k+1)\)\cr
&+& \sum_{\l=0}^{\infty} e^{-t(2\l+a)} \(\sum_{k=0}^{\l-1}\dim(2\l-2k,0) g(2k) + \sum_{k=0}^{\l-1} \dim(2\l-2k-1,0) g(2k+1)\)\cr
&+& \sum_{\l=0}^{\infty} e^{-t(2\l+1+a)} \(\sum_{k=0}^{\l} \dim(2\l-2k+1,0) g(2k) + \sum_{k=0}^{\l-1} \dim(2\l-2k,0) g(2k+1)\)\cr
&+& \sum_{\l=0}^{\infty} e^{-t(2\l+a)} h(2\l) + \sum_{\l=0}^{\infty} e^{-t(2\l+1+a)} h(2\l+1)\label{summavektor}
\een
where 
\bea
f(2k) &=& \sum_{q=0}^k e^{\alpha t 2 q} + \sum_{q=0}^{k-1} e^{\alpha t (2q+2)}\cr
f(2k+1) &=& 2\sum_{q=0}^k e^{\alpha t (2q+1)}\cr
g(2k) &=& 2 \(\sum_{q=0}^k e^{\alpha t (2q+1)}+\sum_{q=0}^{k-1} e^{\alpha t (2q+1)}\)\cr
g(2k+1) &=& \sum_{q=0}^{k+1} e^{\alpha t 2q} + \sum_{q=0}^{k-1} e^{\alpha t (2q+2)} + \sum_{q=0}^k e^{\alpha t 2 q} + \sum_{q=0}^k e^{\alpha t (2q+2)}\cr
h(2\l) &=& 2 \sum_{q=0}^{\l-1} e^{\alpha t (2q+1)}\cr
h(2\l+1) &=& \sum_{q=0}^{\l} e^{\alpha t 2q} + \sum_{q=0}^{\l-1} e^{\alpha t (2q+2)}
\eea
The sum (\ref{summavektor}) can be computed and series expanded for small values of $t$. We used Mathematica for this purpose. From this series expansion we can read off the constant term. It precisely agrees with what we obtained in (\ref{vektor}).

\subsection{Fermion harmonics}
The fermion harmonics form the representation $(n,0,1)$ of $SO(7)$. They appear in the product of one fermion with the scalar harmonics,
\bea
(0,0,1) \otimes (n,0,0) &=& (n,0,1) \oplus (n-1,0,1)
\eea
From the branching rule of the fermion,
\bea
(0,0,1) &\rightarrow & (0,1)_{1/2} \oplus (0,1)_{-1/2} 
\eea
together with the branching rule (\ref{branchscalar}) for the scalar harmonics and the product decomposition
\bea
(0,1)_{\pm 1/2} \otimes (p,0)_q &=& (p,1)_{q\pm 1/2} \oplus (p-1,1)_{q\pm 1/2}
\eea
we can deduce the branching rule for the fermion harmonics,
\bea
(n,0,1) &\rightarrow & \bigoplus_{p=0}^n \bigoplus_{q=0}^p \((n-p,1)_{p+\frac{1}{2}-2q} \oplus (n-p,1)_{p-\frac{1}{2}-2q}\)
\eea
The corresponding refined dimension is 
\bea
f_n(\alpha) &=& \sum_{p=0}^n \sum_{q=0}^p \dim(n-p,1) \(e^{\alpha t |p+1/2-2q|}+e^{\alpha t |p-1/2-2q|}\)
\eea
More explicitly
\bea
f_{2\l}(\alpha) &=& \sum_{k=0}^{\l} 2 \dim(2\l-2k,1) \(\sum_{q=0}^k e^{\alpha t (1/2+2q)} + \sum_{q=0}^{k-1} e^{\alpha t(3/2+2q)}\)\cr
&& + \sum_{k=0}^{\l-1} 2 \dim(2\l-2k-1,1) \sum_{q=0}^k \(e^{\alpha t(3/2+2q)} + e^{\alpha t(1/2+2q)}\)\cr
f_{2\l+1}(\alpha) &=& \sum_{k=0}^{\l} 2 \dim(2\l-2k,1) \(\sum_{q=0}^k e^{\alpha(1/2+2q)} + \sum_{q=0}^{k-1} e^{\alpha t(3/2+2q)}\)\cr
&& + \sum_{k=0}^{\l} 2\dim(2\l-2k+1,1) \sum_{q=0}^k \(e^{\alpha t (3/2+2q)} + e^{\alpha t (1/2+2q)}\)
\eea
The heat kernel is
\bea
K_{\alpha}(t) = \sum_{n=0}^{\infty} f_n(\alpha) e^{-t(n+3)} = \sum_{\l=0}^{\infty} \(f_{2\l}(\alpha) e^{-t(2\l+3)} + f_{2\l+1}(\alpha) e^{-t(2\l+4)}\)
\eea
This sum can be evaluated and then be series expanded for small values for $t$. The result is
\bea
K_{1-\gamma}(t) &=& \frac{8}{\gamma t^6} - \frac{5+\gamma^2}{3\gamma t^4} + \frac{135 + 50 \gamma^2 + 7 \gamma^4}{720 \gamma t^2} -\frac{367}{24196\gamma} - \frac{\gamma}{128} - \frac{7\gamma^3}{3456} - \frac{31\gamma^5}{120960} + ...
\eea
that is a refinement of 
\bea
K_{1}(t) &=& \frac{8}{t^6} - \frac{2}{t^4} + \frac{4}{15 t^2} - \frac{191}{7560} + ...
\eea
that can be computed directly by using 
\bea
K_1(t) &=& \sum_{n=0}^{\infty} f_n(1-\gamma) e^{-t(n+3)}\cr
f_n &=& \frac{1}{15} (n+1)(n+2)(n+3)(n+4)(n+5)
\eea

\section{Supersymmetric conical deformation}\label{susyresults}
The conformal Killing spinor on $S^6$ picks up a phase factor (\ref{phase}) on $S^6_q$ as we go around the circle once by letting $\tau \rightarrow \tau + 2\pi q$. There is no way we can modify the Killing spinor solution since $S^6_q$ is locally the same as $S^6$ and hence the Killing spinor equations on $S^6_q$ and $S^6$ are locally the same, and so their solutions are locally the same. The only candidate solution is the Killing spinor on $S^6$, possibly up to a choice of spin structure, which is global data, but on $S^6$ there is only one spin structure, so there is no such freedom available. However, for the M5 brane, the supersymmetry parameter has an $SO(5)$ R-spinor index, so we can preserve supersymmetry by turning on the background R-gauge field so that it cancels this phase factor for some of the spinor components. In that way we can preserve some amount of supersymmetry on $S^6_q$. 

For a field this amounts to modifying the covariant derivative as $D_{\tau} \rightarrow D_{\tau} + ik(q-1)$ where $k$ is the R-charge. The R-symmetry group is $SO(5)$ and has two Cartan generators $R_1$ and $R_2$. The five scalars form a vector of $SO(5)$. We may form two complex scalars with unit R-charges $(1,0)$ and $(0,1)$ under $(R_1,R_2)$. The complex conjegute scalars carry R-charges $(-1,0)$ and $(0,-1)$. All five scalars have conformal mass. For the complex scalars, the modified half heat kernels are therefore obtained by shifting the $U(1)_q$ quantum number $m \rightarrow m \mp (q-1)$, which amounts to shifting $\gamma m \rightarrow \gamma m \pm (\gamma -1)$, 
\bea
K^{\pm}_a(t) &=& \sum_{n=0}^{\infty} \(d_{n,0} + 2 d_{n,m\neq 0} \sum_{m=1}^{\infty} e^{- t \(\gamma m \pm (\gamma-1)\)}\) e^{-t(n+a)}
\eea
It turns out that with the shift, $K_2^{\pm}(t)$ is no longer equal to $K_3^{\pm}(t)$. We thus need to compute the sum of these and then divide by two, $\(K_2^{\pm}(t) + K_3^{\pm}(t)\)/2$ to get the full half heat kernel. The resulting conformal anomalies that  we get this way are
\bea
a_6^+ &=& \frac{1}{15120 q^5} - \frac{1}{864 q^3} - \frac{19 q}{6048} + \frac{1}{180}\cr
a_6^- &=& \frac{253}{15120 q^5} - \frac{1}{12 q^4} + \frac{119}{864 q^3} - \frac{1}{12 q^2} - \frac{19 q}{6048} + \frac{1}{60}
\eea
It is notable that the Casimir energy terms are identical for $a_6^+$ and $a_6^-$ while all the other terms are different. While this asymmetry between positive and negative R-charges for the scalars is mysterious to us, it was noted already in \cite{Nian:2015xky} in the context of Renyi entropies. From the above, we can compute the corresponding Renyi entropies with the result
\bea
S^+ - S &=& \frac{1}{60 q^5} - \frac{1}{15 q^4} + \frac{17}{240 q^3} - \frac{1}{80 q^2} - \frac{1}{80 q} + \frac{1}{240}\cr
S^- - S &=& - \frac{1}{720 q^3} - \frac{1}{720 q^2} - \frac{1}{720 q} + \frac{1}{240}
\eea
where 
\bea
S &=& \frac{1}{15120 q^5} + \frac{1}{15120 q^4} + \frac{1}{3360 q^3} + \frac{1}{3360 q^2} + \frac{1}{3360 q} + \frac{1}{3360}
\eea
is the Renyi entropy computed from an R-neutral conformal scalar (\ref{ConfScal}). A general formula for $S^{\mu} - S = \Delta S^{s}(\mu)$ was presented in \cite{Nian:2015xky} for a conformal scalar (indicated by the superscipt $s$) with R-charge $\mu$. We find the following agreements with their general formula,  
\bea
S^+ - S &=& \frac{1}{2} \Delta S^s(1-q)\cr
S^- - S &=& \frac{1}{2} \Delta S^s(q-1)
\eea
We should expect the factor of $1/2$ since we have a complex scalar that corresponds to two real scalars. We get an extra minus signs due to the overall minus sign in the relation $q-1=-(\gamma-1)/\gamma$

Let us next consider the fermions. It is easy to see that eigenvalues of the Dirac operator on $S^6_q$ are $n+\gamma m + 3$ from our computation in appendix \ref{Fer}. Thus it shall be possible to recast the heat kernel in the form
\ben
K^0_{1-\gamma}(t) &=& \sum_{n=0}^{\infty} \sum_{m=0}^{\infty} d^F_{n,m} e^{-t (n+\gamma m+3)}\label{expression0}
\een
although we have not done it. Specifically, we have not obtained the degeneracies $d_{n,m}^F$ but used instead a representation theory method where these are implicit. Using this method, we obtained the following heat kernel 
\ben
K^0_{1-\gamma}(t) = \sum_{n=0}^{\infty} f_n(1-\gamma) e^{-t(n+3)}\label{expression}
\een
We would now like to include the coupling to an R-gauge field. We do that by shifting the quantum number $m \rightarrow m \mp (q-1)/2$. But $m$ does not appear explicitly in our expression (\ref{expression}). It does appear in (\ref{expression0}), so let us start there and make the replacement $n+\gamma m \pm (\gamma-1)/2$. That gives our deformed heat kernels,
\ben
K^{\pm}_{1-\gamma}(t) &=& \sum_{n=0}^{\infty} \sum_{m=0}^{\infty} d^F_{n,m} e^{-t (n+\gamma m+3 \pm (\gamma-1)/2)}\label{expression0}
\een
Now this deformation just amounts to multiplication by the overall factor $e^{\mp t(\gamma-1)/2}$ and so we can immediately conclude that in the representation theory expression we get the modified heat kernel as
\bea
K^\pm_{\alpha}(t) = \sum_{n=0}^{\infty} f_n(1-\gamma) e^{-t(n+3\pm (\gamma-1)/2)}
\eea
This deformed heat kernel gives the result in equation (\ref{nice}) for $\t a_6^F$. We get the same result for $\t a_6^F$ for both the plus and the minus signs. It has the structure that we expect of the supersymmetric conformal anomaly. It gives the supersymmetric Casimir energy, as the term linear in $q = 1/\gamma$, and from the $1/q^5$ term we may read off $4$ degrees of freedom (so that to get the $8$ degrees of freedom of the 6d (2,0) tensor multiplet, we shall multiply $\t a_6^F$ by $2$). 

In \cite{Nian:2015xky} the supersymmetric Renyi entropy was obtained on $S^1_q \times \H^5$. It is given by 
\bea
S^{f,SUSY} &=& S^f + \Delta S^f(\mu)
\eea
where, if we insert $\mu = \pm (q-1)/2$, 
\ben
\Delta S^f_q &=& - \frac{1}{1920 q^5} - \frac{1}{1920 q^4} - \frac{1}{288 q^3} - \frac{1}{288 q^2} - \frac{13}{1152 q} + \frac{37}{1920}\label{Delt}
\een
and $\Delta S^f(\mu) = \Delta S^f(-\mu)$. From our results for $a_6^F$ and $\t a_6^F$ in (\ref{aF}) and (\ref{nice}), we can compute the corresponding Renyi entropies,
\bea
S_q &=& \frac{a_6^F(q) - q a_6^F(1)}{1-q}\cr
\t S_q &=& \frac{\t a_6^F(q) - q \t a_6^F(1)}{1-q}
\eea
We get
\bea
S_q &=& -\frac{31}{120960 q^5} - \frac{31}{120960 q^4} - \frac{23}{10080 q^3} - \frac{23}{10080 q^2} - \frac{407}{40320 q} - \frac{407}{40320}\cr
\t S_q &=& \frac{1}{3780 q^5} + \frac{1}{3780 q^4} + \frac{1}{840 q^3} + \frac{1}{840 q^2} + \frac{1}{840 q} - \frac{37}{1260}
\eea
Their difference is in agreement with \cite{Nian:2015xky},
\bea
\t S^f_q - S^f_q &=& \Delta S^f_q
\eea
with $\Delta S^f_q$ given by (\ref{Delt}).

\section{Discussion}\label{disc}
We have shown that conformal anomalies computed on $S^6_q$ agree with corresponding results on $S^1_q \times \H^5$, up to a constant term and to a linear term in $q$. The computation that has been done on $S^1_q \times \H^5$ is more subtle than our computation on $S^6_q$ since $S^1_q\times \H^5$ is noncompact and the spectrum is continuous. There appears a divergence for the zero mode along $S^1_q$ that one needs to regularize. The way this was done in \cite{Casini:2010kt} was to remove that zero mode by subtracting the infinite $q$ asymptotic part, which amounts to subtracting the infinite $q$ asymptotic from our conformal anomalies. Concretely that amounts to multiplying the conformal anomalies by the inverse leading power that generically appears to be $q$ and take the limit $q\rightarrow \infty$. The only term that survives this limit is the asymptotic term that was linear in $q$ before multiplying the whole anomaly by $q^{-1}$. So it is precisely this linear term that gets subtracted by this regularization. Thus we can explain the missing Casimir energies as being due to a careless regularization on the $S^1_q \times \H^5$ side and by using a better way to regularize we should expect to recover the linear term on the $S^1_q \times \H^5$ side as well. For the constant shift associated with the tensor gauge field \cite{Nian:2015xky}, this may arise from missing to take into account boundary degrees of freedom that live on the asymptotic boundary $S^1 \times S^4$ of $S^1 \times \H^5$. A corresponding computation for a Maxwell field on $S^1 \times \H^3$ \cite{Huang:2014pfa} has shown that such a constant shift associated with the gauge field comes from a contribution from the boundary $S^1 \times S^2$. Presumably this computation will have a straightforward generalization to a two-form gauge field in six dimensions on $S^1\times \H^5$. 

In \cite{Yankielowicz:2017xkf} the Renyi entropy for nonabelian (1,0) SCFT's was obtained on $S^1_q \times \H^5$. One may ask if the nonabelian conformal anomaly can also be obtained on this space.\footnote{General results for the nonabelian 6d (2,0) conformal anomaly on smooth six-manifolds have been obtained in \cite{Cordova:2015vwa}, \cite{Maxfield:2012aw}.} The relation between the Renyi entropy and conformal anomaly as given in (\ref{Renyi}) can be at least partially inverted. If the conformal anomaly is on the form\footnote{The leading term $\sim q^{-5}$ may be lowered by supersymmetric cancelation.}
\bea
a_6 &=& c_5 q^{-5} + c_4 q^{-4} + \cdots + c_0 
\eea
where we deleted the linear term in $q$, , then the corresponding Renyi entropy becomes
\bea
S &=& c_5 q^{-5} + (c_4+c_5) q^{-4} + ... + (c_1+...+c_5) q^{-1} + c_0+...+c_5
\eea
Hence by knowing the coefficients in the nonabelian Renyi entropy of some (1,0) SCFT, as were obtained in \cite{Yankielowicz:2017xkf}, we can go back and obtain the conformal anomaly $a(q)$, at least up to the linear term in $q$. That is, up to the knowledge of the nonabelian Casimir energy.

In \cite{Yankielowicz:2017xkf} it was argued that the supersymmetric Casimir energy (possibly rescaled by some factor) can be obtained from the leading term in the supersymmetric Renyi entropy for (1,0) supermultiplets as 
\ben
\lim_{\gamma\rightarrow \infty} \frac{S_\gamma}{\gamma^3} &=& \frac{1}{192}(\alpha-4\beta+16\gamma)\label{casi}
\een
Here $\alpha,\beta,\gamma,\delta$ are the coefficients in the eight-form anomaly polynomial. For the abelian $(1,0)$ tensor multiplet and hypermultiplet these anomaly polynomials were obtained in \cite{Bak:2016vpi}. From that we can read off the coefficients for the abelian $(1,0)$ tensor multiplet as $
\alpha = 1, \beta = 1/2, \gamma = 7/240, \delta = -1/60$ that gives us
\bea
\frac{1}{192}(\alpha-4\beta+16\gamma) &=& - \frac{1}{360}
\eea
The Renyi entropy for the abelian tensor multiplet is computed from 
\bea
a_6^{tensor} = a_6^S+a_6^B-\t a_6^F = \frac{1}{360 q^3} + \frac{1}{12 q} + \frac{11 q}{120} + \frac{3}{8}
\eea
with the result 
\bea
S^{tensor} &=& \frac{1}{360 q^3} + \frac{1}{360 q^2} + \frac{31}{360 q} + \frac{83}{180}
\eea
Thus by looking at the leading term in the Renyi entropy, we confirm that the same number $1/360$ (up to a convention dependent minus sign) appears on both sides in the relation (\ref{casi}). On the other hand, the supersymmeric Casimir energy of the tensor multiplet, which may be extracted from the term linear in $q$ in $a_6^{tensor}$, is given by $E^{tensor} = -11/240$, which is in agreement with \cite{Bak:2016vpi}. It does not seem obvious that the two numbers $-11/240$ and $1/360$ would be related. 

For the hyper we look at the combination (the general case where two Cartans of $SO(5)$ are involved was considered in \cite{Nian:2015xky})
\bea
a^{hyper}_6 = 2 a^+_6 + 2 a^S_6 - \t a^F = - \frac{1}{360 q^3} + \frac{1}{24} - \frac{q}{120}
\eea
where $\phi^1+i \phi^2$ carries R-charge $+1$ and $\phi^3$ and $\phi^4$ are R-neutral. It should be noted that there appears two copies of the holomorphic field $\phi^1 + i \phi^2$ that leads to $2 a^+_6$, rather than one holomorphic and one antiholomorphic field (that is, $\phi^1+i\phi^2$ and $\phi^1-i \phi^2$) that would lead to $a^+_6 + a^-_6$. The holomorphic choice of $2a^+_6$ is necessary to have a supersymmetric cancelation of the $q^{-5}$ terms, but we think it is unlikely this can be understood at the classical level. The holomorphy is presumably related to having a chiral theory in euclidean signture where chiral spinors can not be also made real. More rigorously, this might be possible to trace back to holomorphic factorization of a nonchiral quantum theory \cite{Witten:1996hc}. From $a^{hyper}$ we read off the Casimir energy $E^{hyper} = 1/240$ that corresponds to the Casimir energy at the points of enhanced $(2,0)$ supersymmetry, at the values of a mass parameter $m=\pm 1/2$ in \cite{Bak:2016vpi}.

\subsection*{Acknowledgments}
I would like to thank Yang Zhou for explaining his work during the East Asia Joint Workshop on Fields and Strings 2018 at KIAS and Arkady Tseytlin for bringing his works to my attention by correspondence. This work was supported in part by NRF Grant 2017R1A2B4003095.

\appendix

\section{The box operator on scalar, vector and two-form}\label{Box}
We define the box operator on $\mb{R}^{d+1}$ as
\bea
\square = \delta^{IJ} D_I D_J
\eea
The metric in Euclidean and Polar coordinates is 
\bea
ds^2 = dx^I dx^I = dr^2 + r^2 G_{mn}(\theta) d\theta^m d\theta^n
\eea
We have
\bea
x^I = r f^I(\theta)
\eea
The nonvanishing Christoffel symbols are
\bea
\Gamma^r_{mn} &=& - r G_{mn}\cr
\Gamma^m_{nr} &=& \frac{1}{r} \delta^m_n\cr
\Gamma^p_{mn} &=& \t\Gamma^p_{mn}
\eea
The box operator in Polar coordinates is
\bea
\square = D_r D_r + \frac{1}{r^2} G^{mn} D_m D_n
\eea
Acting on a scalar, we have
\bea
\square \phi = D_r D_r \phi + \frac{1}{r^2} G^{mn} D_m D_n \phi
\eea
Let us define 
\bea
V_r &=& \frac{\partial x^I}{\partial r} V_I\cr
V_m &=& \frac{\partial x^I}{\partial \theta^m} V_I
\eea
We now need the following covariant derivatives,
\bea
D_r V_r &=& \partial_r V_r\cr
D_m V_n &=& \t D_m V_n + r G_{mn} V_r
\eea
where 
\bea
\t D_m V_n &=& \partial_m V_n - \t \Gamma^p_{mn} V_n
\eea
Then we get
\bea
\square \phi &=& \frac{1}{r^2} \t\square \phi + \partial_r^2 \phi + \frac{d}{r} \partial_r \phi
\eea
where we define
\bea
\t\square \phi &=& G^{mn} \t D_m \t D_n \phi
\eea
We notice that the same result is obtained from 
\bea
\square \phi &=& \frac{1}{r^d \sqrt{G}} \partial_M \(r^d \sqrt{G} g^{MN} \partial_N \phi\)
\eea

Let us move on to the vector,
\bea
\square v_p &=& D_r D_r v_p + \frac{1}{r^2} G^{mn} D_m D_n v_p
\eea
We put 
\bea
D_r v_p &=& w_{rp}\cr
D_n v_p &=& w_{np}
\eea
just to avoid distraction. Then we have 
\bea
\square v_p &=& D_r w_{rp} + \frac{1}{r^2} G^{mn} D_m w_{np}
\eea
We expand out these covariant derivatives to get
\bea
\square v_p &=& \frac{1}{r^2} G^{mn} \t D_m w_{np} + \frac{d}{r} w_{rp} + \frac{1}{r} \(w_{pr} - w_{rp}\) + \partial_r w_{rp}
\eea
Now we insert 
\bea
w_{pr} &=& - \frac{1}{r} v_p\cr
w_{rp} &=& \partial_r v_p - \frac{1}{r} v_p\cr
w_{np} &=& \t D_n v_p 
\eea
We put $v_r = 0$. We get
\bea
\square v_p &=& \frac{1}{r^2} \t \square v_p + \partial_r^2 v_p + \frac{d-2}{r} \partial_r v_p - \frac{d-1}{r^2} v_p
\eea
This we can also express as
\bea
\square v_p &=& \frac{1}{r^2} \t \square v_p + \(\partial_r + \frac{d-1}{r}\) \(\partial_r - \frac{1}{r}\) v_m - \frac{1}{r^2} v_m
\eea
if we let $\partial_r$ act on everything that stands to the right, so for instance $\partial_r \frac{1}{r} v_m$ shall be interpreted as $\partial_r\(\frac{1}{r} v_m\)$. Then this agrees with the equation (2.2) in \cite{DeNardo:1996kp}. We now notice that 
\bea
\partial_r v_I &=& 0
\eea
implies that 
\bea
\partial_r v_m &=& \frac{1}{r} v_m
\eea
Using this, we finally get 
\bea
\square v_p &=& \frac{1}{r^2} \t \square v_p - \frac{1}{r^2} v_p
\eea

Let us turn to the two-form. We find 
\bea
\square B_{mn} &=& \frac{1}{r^2} \t \square B_{mn} + \partial_r^2 B_{mn} + \frac{d-4}{r} \partial_r B_{mn} - \frac{2(d-2)}{r^2} B_{mn}
\eea
We now notice that 
\bea
\partial_r B_{IJ} &=& 0
\eea
implies that 
\bea
\partial_r B_{mn} &=& \frac{2}{r} B_{mn}
\eea
Using this, we finally get
\bea
\square B_{mn} &=& \frac{1}{r^2} \t \square B_{mn} - \frac{2}{r^2} B_{mn}
\eea

\section{Laplacian on scalar, vector and two-form}\label{Lap}
We have the following Laplacians on $S^d$,
\bea
\t\triangle \phi &=& - \t\square \phi\cr
\t\triangle v_m &=& - \t\square v_m + R_m^n v_n\cr
\t\triangle B_{mn} &=& - \t\square B_{mn} + R_m^p B_{pn} - R_n^p B_{pm} + 2 R^p{}_m{}^q{}_n B_{pq}
\eea

On $S^6$ we get
\bea
R^p{}_m{}^q{}_n &=& \delta^p_n \delta^q_m - g^{pq} g_{mn}\cr
R^m_n &=& 5 \delta^m_n\cr
R &=& 30
\eea
This leads to
\bea
\t\triangle v_m &=& - \t\square v_m + 5 v_m\cr
\t\triangle B_{mn} &=& - \t\square B_{mn} + 8 B_{mn}
\eea

Eigenvalues are
\bea
\lambda_n &=& n(n+5) - 1 + 5 = n(n+5) + 4
\eea
for Laplace on vector harmonics. For two-form harmonics we get
\bea
\lambda_n &=& n(n+5) - 2 + 8 = n(n+5) + 6
\eea

\section{Factorization of the determinant}\label{factorization}
Here we explain the heat kernel method that was used in \cite{Beccaria:2017lcz} when the eigenvalues factorize. If the eigenvalues of some quadratic differential operator $\triangle$ take the form of a product
\bea
\lambda_n &=& \lambda_n^a \lambda_n^b
\eea
then the corresponding determinant should factorize. When the product is a divergent infinite product, the determinant is defined through a zeta function
\bea
\zeta(s) &=& \sum_n \lambda_n^{-s}
\eea
as
\bea
\det \triangle &=& e^{-\zeta'(0)}
\eea
and the factorization means that
\ben
\zeta'(0) &=& \zeta_a'(0) + \zeta_b'(0)\label{etta}
\een
Let us assume the eigenvalues are on the form 
\bea
\lambda_n &=& \frac{\h\lambda_n}{r^2}\cr
\lambda_n^{a,b} &=& \frac{\h\lambda_n^{a,b}}{r}
\eea
where $\h\lambda_n$ and $\h\lambda_n^{a,b}$ are dimensionless\footnote{Subscript $a,b$ is a short notation that means either $a$ or $b$.}. Then this derivative becomes
\bea
\zeta'(0) &=& 2 \ln(r) \zeta(0) + \h\zeta'(0)\cr
\zeta_{a,b}'(0) &=& \ln(r) \zeta_{a,b}(0) + \h\zeta'_{a,b}(0)
\eea
The relation (\ref{etta}) now amounts to the relation
\bea
\zeta(0) &=& \frac{1}{2} \(\zeta_a(0) + \zeta_b(0)\)
\eea
This means that we can compute $\zeta(0)$ from the half heat kernels
\bea
K_{a,b}(t) = \sum_n e^{-t \lambda_n^{a,b}}
\eea
instead of using the original heat kernel 
\bea
K(t) = \sum_n e^{-t \lambda_n}
\eea
which is much harder to compute in our application where the eigenvalues $\lambda$ are quadratic, while $\lambda_{a,b}$ are linear in $n$.

There seems to be no simple relation for the other heat kernel coefficients. For example, for the massless scalar on $S^6$ with eigenvalues $\t\lambda= n(n+5)$ we get the heat kernel
\bea
K(t) &=& \frac{1}{60t^3} + \frac{1}{12t^2} + \frac{1}{5t} + \frac{1139}{3780} + ...
\eea
and the half heat kernels
\bea
K_a(t) = K_b(t) = \frac{2}{t^6} + \frac{5}{t^5} + \frac{6}{t^4} + \frac{55}{12 t^3} + \frac{149}{60 t^2} + \frac{1}{t} + \frac{1139}{3780} + ...
\eea

\section{Explicit solutions for fermion harmonics}\label{Fer}
Let us comment on explicit solutions for the fermion harmonics. We will restrict ourselves to $S^6$ and only briefly mention $S^6_q$ at the end. Let us begin by considering the conformal Killing spinor equation 
The conformal Killing spinor equation is 
\bea
D_M \eps &=& \Gamma_M \eta\cr
D_M \eta &=& - \frac{1}{4r^2} \Gamma_M \eps
\eea
from which follows that 
\bea
D_M \eps_{\pm} &=& \mp \frac{i}{2r} \Gamma_M \eps_{\pm}
\eea
where
\bea
\eps_{\pm} &=& \eps \pm 2 i r \eta
\eea
We also find that 
\bea
\square \eps &=& -\frac{3}{2r^2} \eps\cr
\square \eta &=& - \frac{3}{2r^2} \eta
\eea
The most general solution is most easily found using stereographic coordinates on $S^6$ where the metric is 
\bea
ds^2 &=& e^{2\sigma} dx^M dx^N\cr
e^{\sigma} &=& \frac{1}{1+\frac{|x|^2}{4r^2}}\cr
\sigma &=& - \ln\(1+\frac{|x|^2}{4r^2}\)
\eea
Then the solutions are simply obtained from the flat space solution $\eps = \eps_0 + x^M \Gamma^{\h M} \eps_1$ by a conformal map. We get
\bea
\eps_\pm &=& e^{\sigma/2} \(1 \mp \frac{i}{2r} x^M \Gamma^{\h M}\) \eps_{0,\pm}
\eea
where
\bea
\eps_{0,\pm} &=& \eps_0 \pm 2 i r \eps_1
\eea
and $\Gamma^{\h M}$ are the flat space gamma matrices obeying $\{\Gamma^{\h M},\Gamma^{\h N}\} = 2 \delta^{\h M \h N}$. 

We now define two scalars out of one spinor as
\bea
\phi_{\pm} &=& \bar\eps_{\pm}\psi 
\eea
where the bar means hermitian conjugate. We then get
\bea
\square \phi_{\pm} &=& - \frac{3}{2} \phi_{\pm} + \bar\eps_{\pm} \square \psi \pm i \bar\eps_{\pm} \Gamma^M D_M \psi
\eea
Let us now assume that 
\bea
\Gamma^M D_M \psi &=& i \mu \psi
\eea
Then also we get
\bea
\square \psi &=& \(-\mu^2+\frac{15}{2}\)\psi
\eea
and so we get
\bea
\square \phi_{\pm} &=& - \(\mu^2 \pm \mu - 6\) \phi_{\pm}
\eea
By demanding that $\phi_{\pm}$ are scalar harmonics with eigenvalues 
\bea
\square \phi_{\pm} &=& - \(n^2 + 5 n\) \phi_{\pm}
\eea
we get the equations
\bea
\mu^2 \pm \mu - 6 &=& n^2 + 5 n
\eea
which have the solutions
\bea
\mu &=& \Bigg\{\begin{array}{c}
\mp (n+3)\\
\pm (n+2)
\end{array}
\eea
Thus we see that $\phi_+$ is composed of fermions with eigenvalues $-(n+3)$ and $n+2 = (n-1) + 3$. But as both $\phi_+$ and $\phi_-$ have the same eigenvalue, they must be joined together into one spherical harmonic multiplet with that eigenvalue ($n^2 + 5 n$), which means that we have a decomposition of one scalar harmonic into two fermionic harmonics, thus realizing the group theory decomposition
\bea
(0,0,1) \otimes (n,0,0) &=& (n,0,1) \oplus (n-1,0,1)
\eea

We would now like to solve
\bea
\Gamma^M D_M \psi &=& i \mu \psi
\eea
We make the ansatz
\bea
\psi_{\pm} &=& a \eps_{\pm} \phi + b \Gamma^M \eps_{\pm} D_M \phi
\eea
Then we get
\bea
\Gamma^M D_M \psi_{\pm} &=& \eps_{\pm} \(\mp 3 i a + b \square\) \phi + \Gamma^M \eps_{\pm} \(a\pm 2 i b\) D_M \phi
\eea
and so the eigenvalue problem reduces to 
\bea
\(\begin{array}{cc}
\mp 3 i & \square\\
1 & \pm 2 i 
\end{array}\) \(\begin{array}{c}
a\\
b
\end{array}\) &=& i \mu \(\begin{array}{c}
a\\
b
\end{array}\)
\eea
The secular equation becomes
\bea
\(\mp 3 - \mu\) \(\mp 2 + \mu\) - \square &=& 0
\eea
We take 
\bea
\mu &=& \Bigg\{\begin{array}{c}
\mp (n+3)\\
\pm (n+2)
\end{array}
\eea
to get $\square = n^2 + 5 n$.

\section{Killing spinor on $S^1_q$}\label{S1}
Let us begin with $\mb{R}^2$ where we may use polar coordinates
\bea
x &=& R \cos \tau\cr
y &=& R \sin \tau
\eea
The metric is 
\bea
ds^2 = dR^2 + R^2 d\tau^2
\eea
The vielbein is 
\bea
E^1 &=& dx\cr
E^2 &=& dy 
\eea
in Cartesian coordinates, and 
\bea
e^{\h R} &=& dR\cr 
e^{\h \tau} &=& R d\tau
\eea
in polar coordinates. We have the following $SO(2)$ tangent space transition when going between these vielbeins
\bea
\(\begin{array}{c}
E^1\\
E^2
\end{array}\)
&=& \(\begin{array}{cc}
\cos\tau & -\sin\tau\\
\sin\tau & \cos \tau
\end{array}\)
\(\begin{array}{c}
e^{\h R}\\
e^{\h \tau}
\end{array}\)
\eea
Thus the $SO(2)$ transition matrix in the vector representation is 
\bea
g = \(\begin{array}{cc}
\cos\tau & -\sin\tau\\
\sin\tau & \cos \tau
\end{array}\) 
\eea
The corresponding transitition matrix in the spinor representation is 
\bea
g &=& \(\begin{array}{cc}
e^{i\tau/2} & 0\\
0 & e^{-i\tau/2}
\end{array}\)
\eea
and a constant spinor $\eps_0$ with respect to $E^1, E^2$ corresponds to a spinor 
\bea
\eps &=& g^{-1} \eps_0
\eea
with respect to $e^{\h R}, e^{\h \tau}$. This is a Killing spinor on $S^1$ that is antiperiodic,
\bea
\eps(t+2\pi) &=& - \eps(t)
\eea
But we would like to have a periodic Killing spinor. Indeed we can get a periodic Killing spinor if we cover $S^1$ by two overlapping coordinate patches. We may take $0<t<2\pi$ in one patch $U_{\alpha}$, and $-\pi<s<\pi$ in the other patch $U_{\beta}$. On the overlap, these coordinates are related as $s=t-\pi$ and the transition matrices on $U_{\alpha}$ and $U_{\beta}$ are related as $g_{\alpha} = - g_{\beta}$. The Killing spinor is now defined on each patch as 
\bea
\eps_{\alpha} &=& g_{\alpha}^{-1} \eps_0\cr
\eps_{\beta} &=& g_{\beta}^{-1} \eps_0
\eea
and on the overlap, there is a transition 
\bea
\eps_{\alpha} &=& - \eps_{\beta}
\eea
that gives us the periodic Killing spinor if we alternate once from $U_{\alpha}$ to $U_{\beta}$ as we go around $S^1$ once, and it gives us the anti-periodic Killing spinor if we alternate twice from $U_{\alpha}$ to $U_{\beta}$ and then back again to $U_{\alpha}$ as we go around $S^1$ once. However, it is only the periodic Killing spinor that generalizes to $S^d$ for $d>1$.

Now with the periodic Killing spinor on $S^1$, it is easy to see that this will pick up a phase factor as we go around $S^1_{q}$ by letting $\tau \rightarrow \tau + 2\pi q$,
\ben
\eps &\rightarrow & \(\begin{array}{cc}
e^{i (q-1)/2} & 0\\
0 & e^{-i (q-1)/2}
\end{array}\) \eps\label{phase}
\een

This is easily generalized to $S^6_q$. We introduce the following coordinates on $mb{R}^7$,
\bea
x^1 &=& R \cos \tau\cr
x^2 &=& R \sin \tau\cr
x^m &=& r n^m(\theta^\alpha)
\eea
where $\theta^i$ are polar coordinates on $S^4$ and $n^m n^m = 1$ for $m=1,...,5$. The metric on $\mb{R}^7$ is 
\bea
ds^2 &=& dR^2 + R^2 d\tau^2 + dr^2 + r^2 d\Omega_4
\eea
and $S^6$ is characterized by the constraint $R^2 + r^2 = 1$. The vielbein is
\bea
e^{\h R} &=& dR\cr
e^{\h \tau} &=& R d\tau\cr
e^{\h r} &=& dr\cr
e^{\h i} &=& r e^{\h i}_{\alpha} d\theta^{\alpha}
\eea
and is related to the Cartesian vielbein $dx^M$ on $\mb{R}^7$ by the transition matrix
\bea
g &=& g_2 g_4
\eea
where $g_2$ is the transition matrix we computed above on $\mb{R}^2$ and $g_4$ is the transition matrix for $\mb{R}^4$ in polar coordinates, which commutes with $g_2$. One class of conformal Killing spinors on $S^6$ is now obtained from the constant spinor on $\mb{R}^7$ by transformig it by a local Lorentz transformation to the above coordintes,
\bea
\eps = g^{-1}_4 g_2^{-1} \eps_0 = g_2^{-1} g_4^{-1} \eps_0
\eea
and again we see that we pick up the phase factor (\ref{phase}) as we go around $S^6_q$ once, along the $\tau$ direction.


\end{document}